\documentclass[aps,prx,preprint,superscriptaddress]{revtex4-2}
\usepackage{amsmath}
\usepackage{amssymb}
\usepackage{epsfig}
\usepackage{epstopdf}
\usepackage[usenames]{color}
\usepackage[normalem]{ulem}
\usepackage{lineno}
\usepackage{xcolor}
\graphicspath{ {./images/}}

\usepackage{hyperref}
\hypersetup{colorlinks = true, citecolor=blue, linkcolor=blue, urlcolor=blue}

\usepackage{tabularx}
\newcolumntype{Y}{>{\centering\arraybackslash}X}  
\DeclareUnicodeCharacter{2212}{-}

\begin{document}
\title{Emergent medium-range order in amorphous silicon}

\author{Zhen Zhang}
\email[Corresponding author: ]{zhen.zhang@cdut.edu.cn}
\affiliation{College of Physics, Chengdu University of Technology, Chengdu 610059, China}

\author{Stephen R. Elliott}
\affiliation{Physical and Theoretical Chemistry Laboratory, Department of Chemistry, University of Oxford, Oxford OX1 3QZ, UK}

\author{Yujie Wang}
\affiliation{College of Physics, Chengdu University of Technology, Chengdu 610059, China}

\author{Walter Kob}
\email[Corresponding author: ]{walter.kob@umontpellier.fr}
\affiliation{College of Physics, Chengdu University of Technology, Chengdu 610059, China}
\affiliation{Laboratoire Charles Coulomb (L2C),
University of Montpellier and CNRS, F-34095 Montpellier, France}
\date{\today}

\begin{abstract}   
{Medium-range order (MRO) is an important feature of amorphous materials, but its nature in terms of atomic arrangement has remained elusive. Here, using large-scale machine-learning-driven atomistic simulations of amorphous silicon, we probe the MRO of this archetypal network glass.
By characterizing the structure in three dimensions, we reveal 
that the symmetry of the atomic arrangement around a central atom shows a crossover as a function of distance. At short distances, the structure is determined by local bonding constraints, which evolve from predominantly tetrahedral environments in low-density amorphous silicon to highly coordinated and strongly disordered local environments in very-high-density amorphous silicon. At larger distances, however,
distinct cubic- and octahedral-symmetry patterns emerge in the three-dimensional atomic-density distribution, 
despite the absence of these symmetries at shorter length scales. 
We find that this crossover is accompanied by a quasi-periodic spatial organization of the topological rings in the structure. The characteristic spacing between ring centers matches the length scale of the first sharp diffraction peak (FSDP) in the structure factor, establishing a direct connection between this reciprocal-space signature of MRO and its real-space organization. These results demonstrate that the FSDP, the spatial correlation of the rings, and emergent orientational symmetries are not independent manifestations of MRO, but reflect different aspects of a common structural organization. Our findings provide a unified real-space picture of MRO in amorphous silicon and suggest that intermediate-range structural order in network glasses emerges from the interplay between local bonding constraints and extended atomic arrangements driven by efficient packing.
}
\end{abstract}

\maketitle

\clearpage
\section{Introduction}
Disordered materials, such as liquids and glasses, lack long-range positional order, but on short-to-intermediate length scales, their  atomic structure exhibits various types of ordering that determine the materials' properties~\cite{elliott1990physics,varshneya2019fundamentals,binder_kob_2011}. 
Short-range order (SRO), which refers to the structure on the length scales of the nearest and second nearest neighbor distances ($\lesssim 5$~\AA\ in atomic systems), has been studied extensively, both experimentally and computationally, and correlations between SRO features (e.g., coordination number, bond angle and their derivatives) and some dynamic properties of the systems have been established~\cite{binder_kob_2011,royall2015role,tanaka2019reveal}. 

In contrast, the structural order on intermediate length scales (5--20~\AA\ in atomic systems)~\cite{elliott_medium-range_1991,salmon2005topological}, also known as medium-range order (MRO), is much less understood. This lack of insight 
can be attributed to the fact that so far it has not been possible to obtain a coherent picture of the MRO that rationalizes the features of the various correlation functions that are used to characterize MRO. 
A further issue that prevents progress on this question is the fact that direct experimental access to the atomic configurations of disordered materials is very challenging.  
(We note that recently it has become possible for certain materials, including a-Si, to access directly the atomic positions, albeit with limited spatial resolution and system size~\cite{yang2021determining,liao2026accurate,busch2026limit}.) 

Amorphous silicon (a-Si), a paradigmatic disordered material, has been extensively studied in the past due to its fundamental and technological importance~\cite{voyles2003medium, street1999technology,lin2016mesoporous,liu2025amorphous,lewis2022fifty,treacy2012local,roorda2012comment,birney2018amorphous,powell1989the}. This material is also known to  
undergo pressure-induced structural transitions from a low-density amorphous (LDA) to transient high-density amorphous phases with different densities and structures 
at approximately 12--16~GPa
(depending on temperature, compression rate, and sample preparation) before crystallization occurs~\cite{mcmillan2005density,deringer2018realistic,fan2024nc,durandurdu2001ab,morishita2004high,deb2001pressure,daisenberger2011polyamorphic}. 
Recent simulation studies using machine-learning potentials (MLPs) have investigated the microscopic mechanisms and pathways of the pressurization-induced structural transitions in a-Si, focusing primarily on short-range structural descriptors~\cite{deringer2021nature,fan2024nc}. 
However, the consequences of these transitions for the MRO have so far not been investigated, and hence we lack at present a deeper understanding regarding the implication of  these transitions for the properties of the material.

One well-known hallmark of the presence of MRO is the first sharp diffraction peak (FSDP) in the static structure factor $S(q)$, a feature that is present in many network glass-formers~\cite{price1988intermediate}. Despite the large number of studies that have focused on the FSDP in a-Si, its structural origin has remained controversial~\cite{phillips1981topology,uhlherr1995generation,christie2004structural,shi2019distinct,salmon2019ordering,kohara2021relationship,dahal2021origin,dahal2022extended}.
Since a-Si has a network structure, topological rings have often been used to characterize its MRO, see, e.g., Ref.~\cite{deringer2018realistic}. However, to what extent the spatial organization of such medium-range structural motifs 
is connected to the FSDP has so far remained unclear. Another aspect that has received insufficient attention
is the three-dimensional (3D) structural organization of the network, notably beyond short-range distances. Answering this question requires higher-order correlation functions that are sensitive to angular correlation and are not restricted to short-range distances.

In this work, we address the aforementioned questions regarding the MRO in realistic atomistic models of a-Si, generated by using MLP-driven simulations. By  using structural descriptors that go beyond the standard two-body correlation functions, we reveal how in these glasses MRO features emerge at low and high densities. We find that the various manifestations of the MRO, such as the FSDP in $S(q)$, ring structure, and 3D orientational ordering, originate from a common 
structural organization that crosses over from local-bonding-defined order to packing-driven arrangements on extended length scales. These insights not only deepen our understanding of the MRO in a-Si, but may also help elucidate the MRO in other disordered systems with similar atomic structures.

\section{Methods}
\subsection{Simulation details}
We used a-Si models that were generated using a recently introduced general-purpose Gaussian approximation interatomic potential (GAP-18)~\cite{bartok2018machine}. The validity of the Gaussian approximation approach for fitting ML-based interatomic potential has been examined in detail in previous works~\cite{bartok2010gaussian,deringer2021gaussian}. Recent studies have shown that this ML potential allows one to reproduce the various experimental observables of a-Si, including calorimetric enthalpies, $^{29}$Si solid-state NMR shifts, structure factor, and multiple structural/phase transitions under compression~\cite{deringer2018realistic,bernstein2019quantifying,deringer2021nature}. 

The a-Si models contained 100,000 atoms, corresponding to a cubic simulation box of $\approx 13$~nm at ambient conditions. We note that such a large sample size is necessary to resolve intermediate/long-range correlations and allow for good statistics of the structural analysis.
The specific procedures used for generating the a-Si models that we study in this work are as follows. Firstly, the ambient-pressure a-Si model was produced by a carefully designed melt-quench procedure starting from 1500~K and using three cooling rates that depend on temperature. Subsequently, the sample was hydrostatically compressed at 500~K up to 20~GPa over 200~ps at a constant rate. (We refer the reader to Ref.~\cite{deringer2021nature} for details). In order to discern the influence of temperature on the results, we performed additional pressurization runs for the a-Si model at 100~K using the same compression rate as at 500~K. All simulations were carried out using the Large-scale  Atomic/Molecular Massively Parallel Simulator software (LAMMPS)~\cite{plimpton_fast_1995} with a time step of 1.0~fs. 

\subsection{Structural quantities for characterizing the MRO}
\textbf{Radial distribution function and structure factor.} The most common standard quantity to characterize the structure of disordered systems is the pairwise radial distribution function, which is given by
\begin{equation}
g(\vec{r})=\frac{V}{N(N-1)}\sum_i\sum_{j\ne i}
\left\langle \delta(\vec{r}+\vec{r}_i-\vec{r}_j)\right\rangle,
\label{eq_gr}
\end{equation}
Here $\langle \cdot \rangle$ denotes the thermal average, $V$ is the volume of the simulation box, $N$ is the total number of atoms, and $\delta$ is the delta-function. 
A further standard quantity to probe the structure is the structure factor, which is defined as 
\begin{equation}
    S(\vec{q})=\frac{1}{N}\sum_{j=1}^N\sum_{l=1}^N \left\langle \exp[-i\vec{q}\cdot(\vec{r}_j-\vec{r}_l)]\right\rangle \quad .
    \label{eq_sq}
\end{equation}
For isotropic systems, such as the glasses considered here, $g(\vec r)$ and $S(\vec q)$ are isotropic and thus the arguments become $r=|\vec r|$ and $q=|\vec q|$, respectively. 

Based on $g(r)$, the coordination number $Z$ of a given atom can be determined by counting its neighboring atoms up to the first minimum distance of $g(r)$, which is chosen to be 2.85~\AA\ for the studied a-Si models. Once the coordination environment is determined, one can further calculate the bond angle $\theta$ between  the central atom and a pair of its nearest neighbors.\\

\textbf{Four-point correlation functions.} We first construct a local coordinate system using a Si atom (particle 1) and any two of its nearest neighbor Si atoms (particles 2 and 3): We define the position of particle 1 as the origin, the direction from particles 1 to 2 as the $z-$axis, and the plane containing the three particles as the $z-x$ plane. This local reference frame allows one to define a spherical coordinate system $(\theta,\phi,r)$, and to measure the probability of finding any other particle at a given point in space, i.e., to measure a structural  
four-point correlation function (see Ref.~\cite{zhang2020pnas} for details). Note that this coordinate system can be defined for all triplets of neighboring particles, and these spatial density distributions can be averaged to improve the statistics. Since this coordinate system is adapted to the local arrangement of the three particles, it allows one to detect angular correlations in space that are not visible in $g(r)$ or other standard structural observables. 

The so-obtained spatial distribution of the particles, $\rho(\theta,\phi,r)$, can be analyzed in a quantitative manner by decomposing it into spherical harmonics, $Y_l^m$,

\begin{equation}
\rho(\theta,\phi,r) = \sum_{l=0}^\infty \sum_{m=-l}^{l}\rho_l^m(r) Y_l^m(\theta,\phi),
\label{eq7}
\end{equation}

\noindent
where the expansion coefficients $\rho_l^m$ are given by

\begin{equation}
\rho_l^m (r)=\int_0^{2\pi} d\phi \int_0^\pi d\theta \sin \theta \rho(\theta,\phi,r) Y_l^{m*}(\theta,\phi) \quad.
\label{eq8}
\end{equation}

Here, $Y_l^{m*}$ is the complex conjugate of the spherical-harmonic function of degree $l$ and order $m$~\cite{press2007numerical}.  In practice, the integral was carried out by sampling the integrand at a given radius $r$ over up to $10^8$ points using a shell of width 0.5~\AA. The strength of the 3D order can then be characterized by the square root of the angular power spectrum,  

\begin{equation} 
S_\rho(l,r)=\Bigl[(2l+1)^{-1}\sum_{m=-l}^{l}|\rho_l^m(r)|^2\Bigr]^{1/2}.
\label{eq9}
\end{equation}

\textbf{Topological rings.} We define the ring size $n$ as the number of Si atoms in the smallest closed loop of Si atoms, the so-called primitive rings~\cite{yuan2002ring}, i.e., none of the identified rings can be decomposed into two smaller rings. The spatial extent of a ring can be characterized by its radius of gyration, $R_{\rm g}$, given by
\begin{equation}
R_{\rm g}^2 = \frac{1}{n}\sum_{i=1}^{n} ({\vec{r}_i} - \vec{R}_{\rm C})^2 \quad,
\end{equation}
where $\vec{R}_{\rm C}$ is the location of the geometrical center of the ring.

\section{Results}
\subsection{Macroscopic response and the corresponding change of local structure}

\begin{figure*}[ht]
\center
    \includegraphics[width=0.95\textwidth]{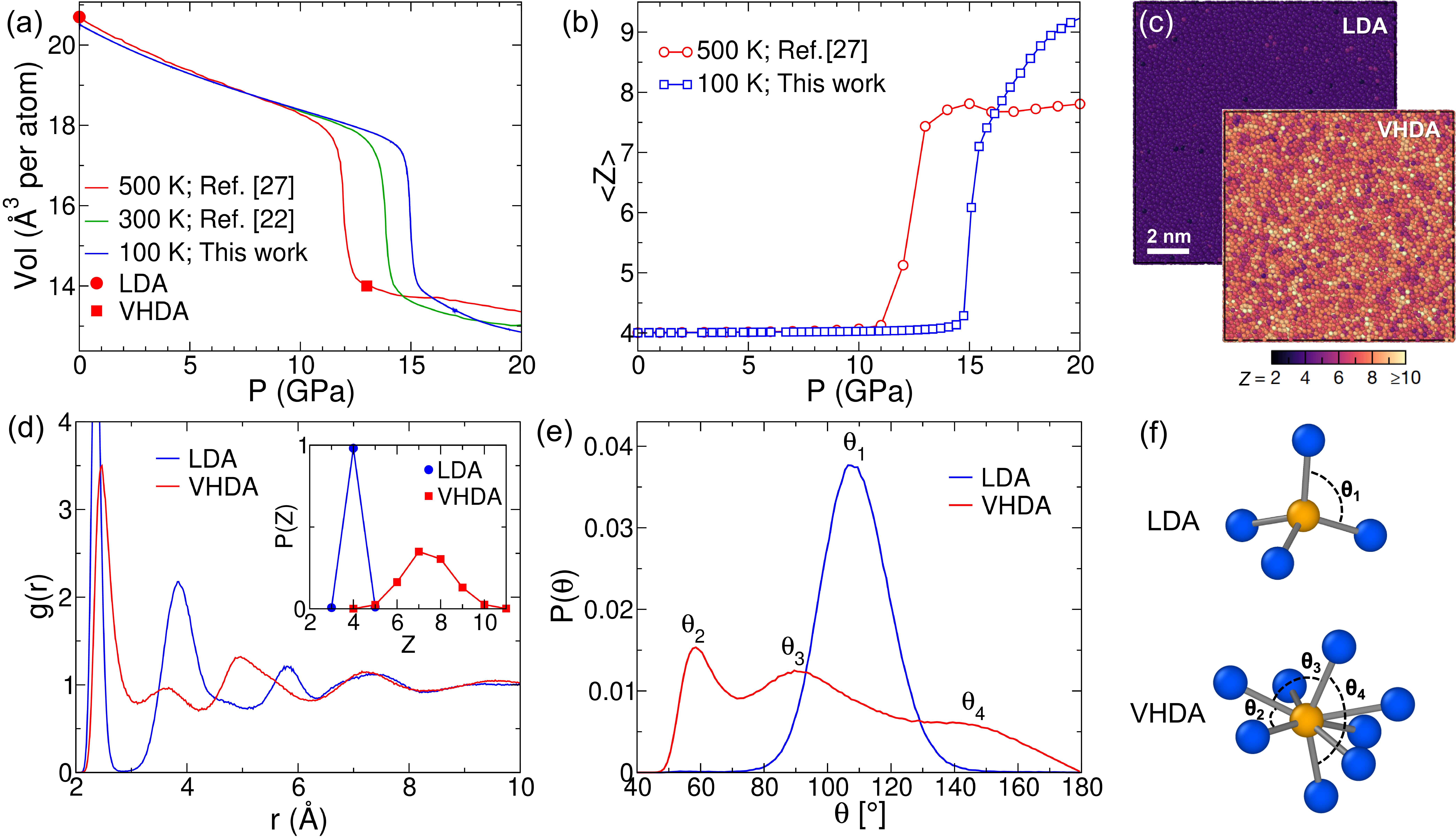}
    \caption{(a) Effect of temperature on the equation of state for a-Si. The red circle and square mark, respectively, the LDA and VHDA states. The data from Deringer et al.~\cite{deringer2021nature} and Fan and Tanaka~\cite{fan2024nc} are included for comparison (all used the same compression protocol).
    (b) Mean coordination number. 
    (c) Structural snapshots for LDA and VHDA phases (reproduced from Ref.~\cite{deringer2021nature}). Color coding indicates the coordination number, $Z$. 
    (d) Radial distribution function $g(r)$. The inset shows the probability distribution that an atom has coordination number $Z$. The mean values are 4.0 and 7.4 for LDA and VHDA, respectively. (e) Probability distribution of bond angles, $\theta$. The characteristic angles are $\theta_1\approx 110^\circ$ for LDA, and $\theta_2\approx  60^\circ$, $\theta_3\approx 90^\circ$, $\theta_4\approx 150^\circ$ for VHDA. (f) Snapshots showing representative local structural motifs with $Z=4$ and $Z=8$ for LDA and VHDA, respectively. The central Si and its nearest neighbors are colored in orange and blue, respectively. The mentioned characteristic angles are also labeled.
    Coordination (and bonds) are determined when the distance between two atoms is smaller than 2.85~\AA, 
    i.e., a distance close to the first minimum in $g(r)$ of the two phases, consistent with the choice of previous studies for the same structural models~\cite{deringer2021nature,fan2024nc}. 
    }
    \label{fig_eos}
\end{figure*}

We first present in Fig.~\ref{fig_eos}(a) the equation of state of the a-Si system, at three different temperatures, which characterizes the macroscopic response of the sample to compression. One recognizes immediately that temperature affects the amorphous-amorphous transition, visible as an abrupt decrease of the volume, mainly in a  quantitative  manner: The sharp structural transition at intermediate pressures persists  but is shifted to higher pressures if the temperature is decreased. Figure~\ref{fig_eos}(b) demonstrates that before the sharp transition, e.g., up to 11~GPa and 15~GPa  for  500~K  and 100~K, respectively, the  average coordination number $\langle Z \rangle$ remains constant, i.e., the network connectivity remains essentially unchanged. To probe the structural difference between the two amorphous phases, we will, in the following, focus  on the two 500~K homogeneous amorphous phases at zero~GPa, i.e., a low-density amorphous (LDA) phase and  at  13~GPa, i.e., the very-high-density amorphous (VHDA) phase~\cite{deringer2021nature}, marked in panel (a) by a red circle and square, respectively.

It should be noted that the VHDA phase is different from the high-density amorphous (HDA) phase often considered in earlier studies (e.g., Ref.~\cite{fan2024nc}), which usually refers to amorphous phases with $Z=5-6$, while the VHDA has $\langle Z \rangle \approx 7.5$. These studies have demonstrated that also in terms of crystallization path, the structure at 13~GPa can be considered as a phase that is distinct from HDA~\cite{fan2024nc}. (We note that, in contrast to Ref.~\cite{fan2024nc}, a homogeneous HDA phase is avoided in our simulations, owing to the compression protocol used.)
We also mention that in Fig.~\ref{fig_eos}(b) the flattening of the 500~K curve at $P\ge14$~GPa is due to crystallization~\cite{deringer2021nature}, while at 100~K the system stays amorphous even up to 20~GPa. This implies that the system is trapped in a very-high-density amorphous phase (with a specific volume of 12.8~\AA$^3$/atom at 20~GPa), having a density that is even higher than the crystalline counterpart with a simple hexagonal structure.
This observation indicates that, due to the low temperature, the system resides in a part of configuration space in which the barriers for crystallization are so high that the amorphous phase is very stable. Hence, further densification of the structure results in a continuous increase of $Z$.

Figure~\ref{fig_eos}(c) shows the corresponding atomic configurations of the two phases in the 500~K samples, with the atoms color-coded, based on their coordination number $Z$. 
One sees that the VHDA phase has not only a higher mean value of $Z$ than the LDA phase, but also a much broader distribution of $Z$, indicating that VHDA has a more complex local structure (see also the inset of Fig.~\ref{fig_eos}(d)). This complexity is not easily recognizable in the radial pair distribution function $g(r)$, Fig.~\ref{fig_eos}(d), but it is reflected in the bond-angle distribution, Fig.~\ref{fig_eos}(e): While for LDA this distribution has only one narrow peak at the intra-tetrahedral angle of $\theta_1\approx  110^\circ$, the distribution of bond angles is much broader in VHDA and exhibits three distinct peaks (or shoulders) at the angles of  $\theta_2\approx 60^\circ$, $\theta_3\approx 90^\circ$, and $ \theta_4\approx 150^\circ$.  The atomic arrangements giving rise to these angles can be understood from  the representative local structural motifs in the two phases shown in  Fig.~\ref{fig_eos}(f). Hence, one concludes that the SRO, as seen from the nearest-neighbor structural observables, is relatively uniform  in LDA whereas it is much more diverse in VHDA.

\begin{figure*}[ht]
\center
    \includegraphics[width=\textwidth]{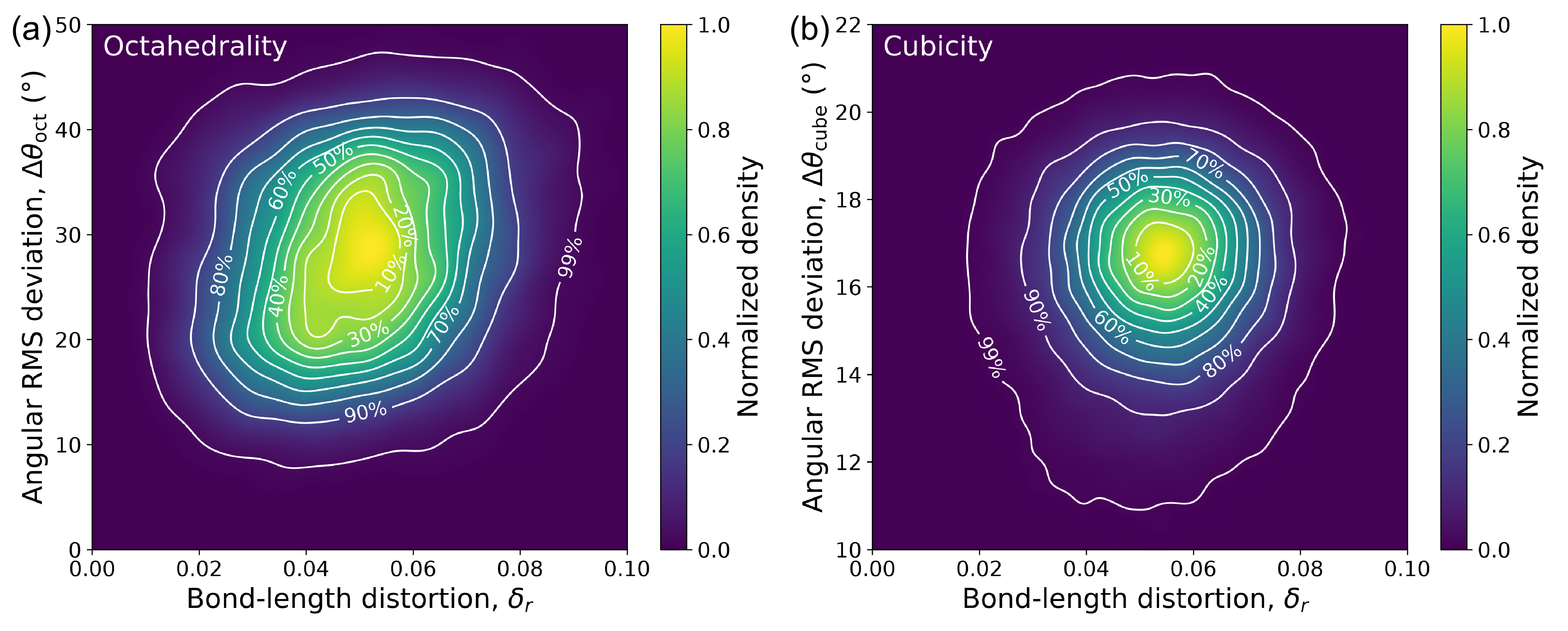}
    \caption{(a)–(b) Two-dimensional density maps characterizing the distortions of the local environments of Si in VHDA with $Z=6$ and $Z=8$ relative to ideal octahedral and cubic geometries, respectively. See the main text for definitions of $\delta_r$ and $\Delta\theta$. Contour lines are labeled by the percentage of data they enclose. 
    }
    \label{fig_SI_regularity-local-struc-vhda}
\end{figure*}

A further important aspect of the SRO is the degree to which the local motifs are distorted. This is particularly relevant for the $Z=6$ and $Z=8$ environments, which are abundant in VHDA and, if sufficiently regular, would give rise to octahedral and cubic symmetries, respectively. Figure~\ref{fig_SI_regularity-local-struc-vhda} shows the density maps of two parameters characterizing the distortion of local environments of Si in VHDA with $Z=6$ and $Z=8$, relative to a regular octahedron and cube, respectively. Specifically, the horizontal axis shows the bond-length distortion, $\delta_r=\sqrt{\langle (r_k-\bar{r})^2\rangle}/\bar{r}$, where $r_k$ is the distance from the central atom to its $k$-th neighbor,
and both $\bar r$ and $\langle\cdots\rangle$ are evaluated locally over
the $Z$ nearest neighbors of that atom.  
The vertical axis represents $\Delta\theta_{\rm oct}$ and $\Delta\theta_{\rm cube}$, defined as the root-mean-squared (RMS) angular deviations from the best ideal octahedral and cubic geometries, respectively. Thus, $\delta_r$  indicates whether the octahedron/cube is radially distorted, whereas $\Delta\theta$ quantifies the degree of angular distortion. One observes that, for both values of $Z$, the angular distortion shows a broad distribution with no clear low-distortion population. This indicates that, despite the large populations of Si  atoms  with $Z=6$ (16\%) and $Z=8$ (30\%) in VHDA, the local motifs do not exhibit pronounced octahedral and cubic symmetries. The emergence of such
symmetries at larger distances, discussed below, is therefore highly nontrivial.

\subsection{Two-point structural correlations}
\begin{figure*}[ht]
\center
     \includegraphics[width=\textwidth]{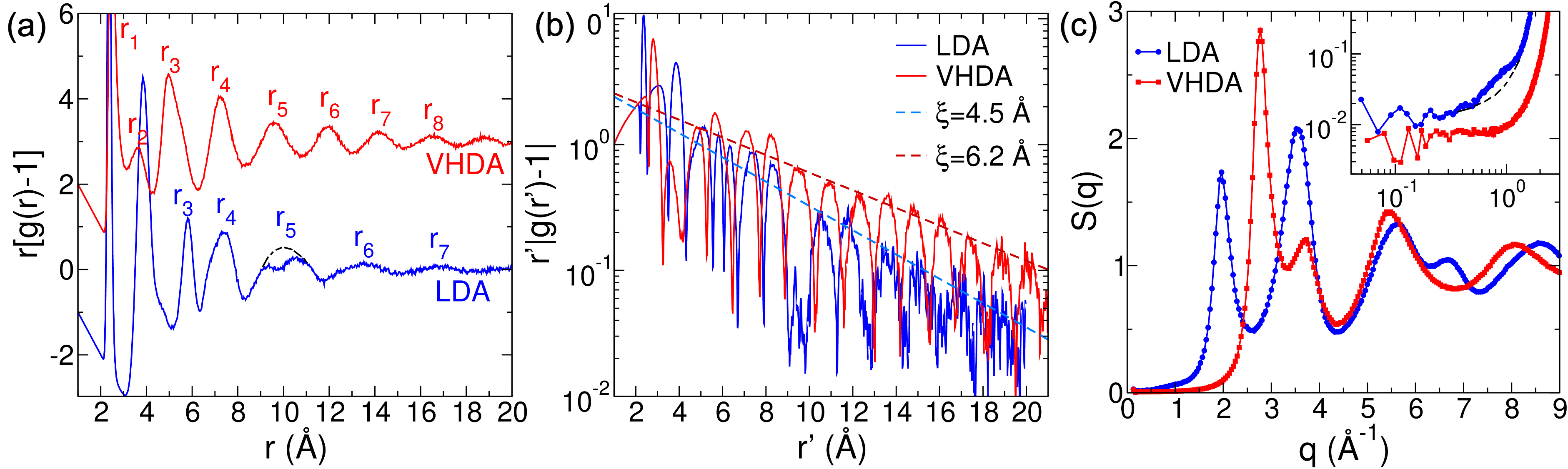}
    \caption{ MRO as revealed from two-point correlation functions. (a) 
    The correlation function $r[g(r)-1]$, motivated by the Ornstein-Zernike relation, for the two systems. The data for VHDA are shifted upward by 3 for clarity. The various peak positions are marked as $r_n$. The dashed line for LDA is a presumed 
    curve, assuming that there is no splitting of the peak at $r\approx 10$~\AA, see main text for details. (b) Logarithmic plot of $r'|g(r')-1|$. Here, the scaled distance $r'=r \cdot L(0)/L(P)$ is used to remove the effect of affine compaction,  where $L(P)$ is the simulation box size at pressure $P$. Fitting of the maxima to an exponential function defines the decay length, $\xi$. (c) Structure factor, $S(q)$, of LDA and VHDA a-Si. The inset shows the low-$q$ region in a double-logarithmic plot. The dashed line in the inset is a presumed baseline to allow better identification of the excess signal in $S(q)$ in LDA at $q \approx 0.7$~\AA$^{-1}$. 
    }
    \label{fig_2point}
\end{figure*}

Next, we explore how these local structural differences propagate to extended length scales. 
To this end, we first consider  the radial pair distribution function, $g(r)$, i.e., the  standard two-point correlation function which shows how the density fluctuations evolve with  distance, $r$.
Fig.~\ref{fig_2point}(a) shows that, compared to LDA, the heights of the peaks (located at $r_n$) in $r[g(r)-1]$ decay more slowly in VHDA, indicating a more pronounced structural correlation in the latter structure. 
This is quantified in  Fig.~\ref{fig_2point}(b) which  shows that, for intermediate and large length scales, the decay of $r'|g(r')-1|$ is described well by an exponential, as expected from the Ornstein-Zernike relation~\cite{ornstein1914influence}. Here $r'=r\cdot L(0)/L(P)$ is a scaled distance that takes into account the size of the box at pressure $P$, $L(P)$.
The length scales $\xi$ extracted from the exponential fits are $\xi \approx 4.5$ and 6.2~\AA\ for LDA and VHDA, respectively, confirming that the structure of VHDA is correlated over larger scales than in LDA. 
One also recognizes from Fig.~\ref{fig_2point}(b) that at intermediate and large distances the signal for LDA is more noisy than the one for VHDA, which indicates that the former structure is more disordered, in accordance with its smaller coherence length.

In addition to the overall decay behavior of $g(r)$,  the correlation functions show two unusual features:  For LDA, one observes in Fig.~\ref{fig_2point}(a) an unexpected double-peak structure, with a minor peak at $r\approx 9.5$~\AA\ and a  larger one at $r\approx 10.5$~\AA. The height of the first peak is close to zero, i.e., $g(r) \approx 1.0$.
(We note that the dip between the two peaks is present at all pressures before the amorphous-amorphous transition, i.e., this is a robust feature of the LDA phase.) 
For the VHDA phase, one finds an unexpected small peak at $r_2$, which also has an intensity close to zero, and which has also been observed in a previous \textit{ab initio} study~\cite{durandurdu2001ab}.
So far, the origin of these features in $g(r)$ for both LDA and VHDA has not been elucidated. 
As we will see below in the context of Fig.~\ref{fig_3d-density}, the analysis of a four-point correlation function permits one to rationalize these peculiar features and to demonstrate that they are related to the occurrence of two competing structures on the corresponding length scale.

The static structure factor $S(q)$, i.e., the space Fourier transform of $g(r)$, is accessible from scattering experiments, and often used as a quantity for probing the MRO. Figure~\ref{fig_2point}(c) shows $S(q)$ for the two systems and from the inset one recognizes that at small wave-vectors the $q$-dependence  in the two phases is very different: At low wave-vectors, both curves flatten and converge to a nonzero constant  value, indicating a normal compressible behavior. 
However, the significantly reduced plateau height at low $q$ for VHDA demonstrates a pronounced suppression of long-wavelength density fluctuations, indicating that VHDA is spatially more homogeneous than LDA, which is consistent with the  larger number of  structural constraints in VHDA and the larger correlation length for the decay of $g(r)$ in panel (b). 
From the inset of Fig.~\ref{fig_2point}(c), one also sees that for LDA there is an excess of intensity (relative to the presumed  baseline dashed curve) at $q\approx  0.7$~\AA$^{-1}$, which corresponds in real space to a distance of around 9~\AA. It is hence tempting to consider this feature in $S(q)$ as the $q$-space manifestation of the splitting of peaks in $g(r)$ of LDA as discussed in the context of panel~(a). 
Independently of the properties at small wave-vectors, we recall that  the characteristic features of the FSDP in $S(q)$  are often used to quantify the MRO. Figure~\ref{fig_2point}(c) shows that, for the LDA sample, the FSDP is at $q_{\mathrm{FSDP}} \approx 2$~\AA$^{-1}$, corresponding in real space to a characteristic length of $\approx 3.1$~\AA. 
In VHDA, the first peak is found to be at $q \approx 2.8$~\AA$^{-1}$, corresponding to a characteristic length $\approx 2.25$~\AA, close to the nearest-neighbor distance. We therefore refer to this feature  in $S(q)$ for VHDA  as the principal diffraction peak rather than an FSDP. The proximity of this characteristic length to the nearest-neighbor length scale suggests that it is predominantly associated with short-range packing. 
In the following, we examine in more detail how the disappearance of the LDA FSDP and the emergence of the VHDA principal peak are related to the reorganization of topological and orientational correlations in real space.

\subsection{Spatial organization of the rings and its connection to the FSDP}

\begin{figure*}[ht]
\center
     \includegraphics[width=\textwidth]{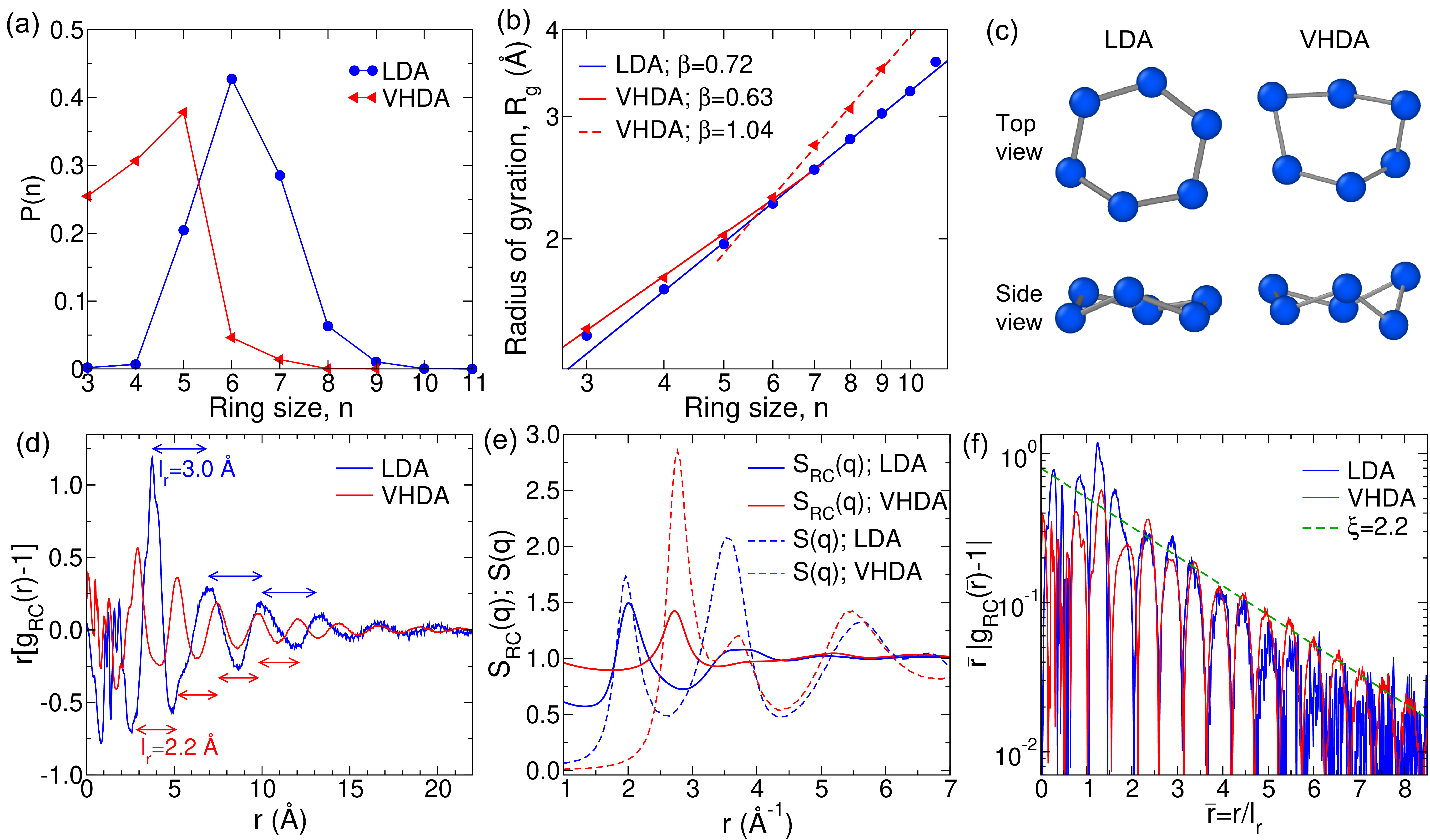}
    \caption{MRO as revealed from the topological ring structure. (a) Probability density distribution of the ring size, $n$. (b) Radius of gyration versus ring size, in a double-logarithmic plot. The blue solid line is a power-law fit $R_g\sim n^\beta$ to the data of LDA, excluding the two points at the two ends. The red solid line and the red dashed line are power-law fits to the data of VHDA for $n\le 6$ and $n\ge 6$, respectively. (c) Typical atomic configurations of six-membered rings in the two phases. (d) Ring center radial distribution function $r[g_{\rm RC}(r)-1]$. The period as seen from the peaks of the curve is marked by the arrows, defining the length scale, $l_r$. (e) Structure factor as obtained from the ring-centers, $S_{\rm RC}(q)$, full lines, compared with the atomic $S(q)$, dashed lines. 
    (f) Semi-logarithmic plot of $\bar{r}|g_{\rm RC}(\bar{r})-1|$ as a function of the scaled distance $\bar{r}=r/l_r$. Fitting of the maxima to an exponential function, the green dashed line, defines the decay length, $\xi$.
    }
    \label{fig_ring-para}
\end{figure*}

We consider the geometrical aspects of the MRO as seen from the characteristics of the rings (see Methods for definition), i.e., the structural units that become relevant beyond the nearest neighbor distance. Figure~\ref{fig_ring-para}(a) shows that the ring-size distribution for LDA peaks at ring size  $n=6$~\cite{deringer2021nature}. In contrast, the VHDA   structure has significantly more  small-sized rings, resulting in a shift of the peak position to $n=5$ and a substantial decrease in the number of large rings. Thus, one can conclude 
that the compaction of the structure is not only caused by an increase of coordination number but also by a strong modification of the ring topology. The geometry of the rings can be characterized via their radius of gyration $R_g(n)$ (see Methods for definition), 
presented in a double-logarithmic plot in Fig.~\ref{fig_ring-para}(b). One sees that for LDA, the increase of $R_g$ with size $n$ can be described by a single power-law with an exponent $\beta=0.72$, i.e., smaller than the value of 1.0, indicating that the rings become increasingly crumpled/elongated with increasing size. 
Interestingly, this exponent is very close to the value found for silica glass, i.e., $\beta \approx 0.75$~\cite{zhang2025pnas}, which hints that, in the two systems, the geometry of the rings is similar since it is dictated by the angular constraints between neighboring bonds. 

In contrast to LDA, one recognizes that the $R_g(n)$ of VHDA shows two power-laws with a crossover at $n=6$, demonstrating that small and large rings scale
differently. For small rings, the exponent, $\beta=0.63$, is somewhat smaller than the value found for LDA, implying a more pronounced change of the geometric shape of the rings with increasing size. This behavior is likely due to the enhanced flexibility of the bond angles, which allows the rings to become more crumpled. Snapshots for typical six-membered rings are shown in Fig.~\ref{fig_ring-para}(c), illustrating that this representative ring structure is more elongated and puckered in VHDA than in LDA.  
At larger ring sizes, one finds instead a power law with an exponent close to the trivial value of 1.0, indicating that the spatial extent of the rings grows nearly linearly with ring size.

Since for a-Si the rings are important structural entities, it is useful to probe how they are arranged with respect to each other.
To this end, we determine their geometric centers, from which we calculate the corresponding center-to-center radial distribution function, $g_{\rm RC}(r)$. Figure~\ref{fig_ring-para}(d) shows this correlator in its reduced form, i.e., $r[g_{\rm RC}(r)-1]$. The peaks at $r<2.5$~\AA\ 
are related to the fact that ring centers are mere geometrical points and hence there is no minimal distance between them, i.e., these peaks are physically not relevant. 
More importantly, one recognizes that, for $r>3$~\AA\, the peaks are spaced in a regular manner and are visible even at large distances. The arrows indicate the presence of a pseudo-period of $l_r\approx3.0$~\AA\ and $\approx$2.2~\AA\ for  LDA and VHDA, respectively. These values coincide with the spatial scale $2\pi/q$ of the FSDP of LDA ($q =2.0$~\AA$^{-1}$) and the principal diffraction peak of VHDA ($q=2.8$~\AA$^{-1}$), respectively (Fig.~\ref{fig_2point}(c)). 
Hence, one can conclude that the location of the first diffraction peak can be read off directly from $g_{\rm RC}(r)$. In other words, the location of the first diffraction peak is given by the pseudo-period corresponding to the distance between ring centers, i.e.,  the rings  are the  relevant  superstructural units that define the MRO. This connection is confirmed by the ring-center structure factor $S_{\rm RC}(q)$, whose dominant peak coincides with the atomic first diffraction peak, see Fig.~\ref{fig_ring-para}(e).

Treating the rings as relevant superstructural units, one can present the corresponding radial distribution function $g_{\rm RC}(r)$ as a function of $\bar{r}=r/l_r$. Figure~\ref{fig_ring-para}(f) shows $\bar{r}[g_{\rm RC}(\bar{r})-1]$ for the two systems and one recognizes that beyond the first two peaks, the two correlators track each other very well. The decay of the peak height is exponential, with a correlation scale of $\xi=2.2$. Hence the correlation length in real units is $\xi \cdot l_r$, which gives 6.6~\AA\ and 4.8~\AA, for LDA and VHDA, respectively. (We note that these length scales measure correlations between ring centers and should be distinguished from the ones extracted from $g(r)$ (Fig.~\ref{fig_2point}(c)) which characterizes atomic-density correlations.)
At intermediate and large length scales, the structure can thus be viewed as a packing of superstructural units, although this analogy should not be interpreted as implying hard-sphere-like packing. In network-forming systems, such as a-Si and silica, the local structural organization is mainly governed by directional bonding and energetic constraints, in contrast to hard-sphere systems in which the packing is determined on all length scales by entropy and excluded volume~\cite{tanaka2019reveal,yuan2021prl}. The present results therefore highlight a crossover from local, bond-defined order at short distances to 
a structure that is determined by the packing of rings at intermediate and large distances.

\subsection{Emergent symmetry beyond local order}

The spatial correlations between ring centers suggest that packing becomes increasingly important at extended length scales. We therefore probe whether the three-dimensional orientational structure also undergoes a crossover from local, bond-defined order to packing-driven organization.
To gain insights into this question, we use a recently proposed four-point correlation function (see Methods for definition) to analyze the 3D structure of the LDA and VHDA a-Si models. 
Since this method is based on a coordinate system that is adapted to local Si-Si-Si linkages, it allows one to detect orientational correlations that are not seen from standard two-point correlation functions. In addition, the radial and density dependence of these large-scale structures may provide clues on the origin of the unusual features seen in $g(r)$ (Fig.~\ref{fig_2point}).

\begin{figure*}[ht]
\center
     \includegraphics[width=\textwidth]{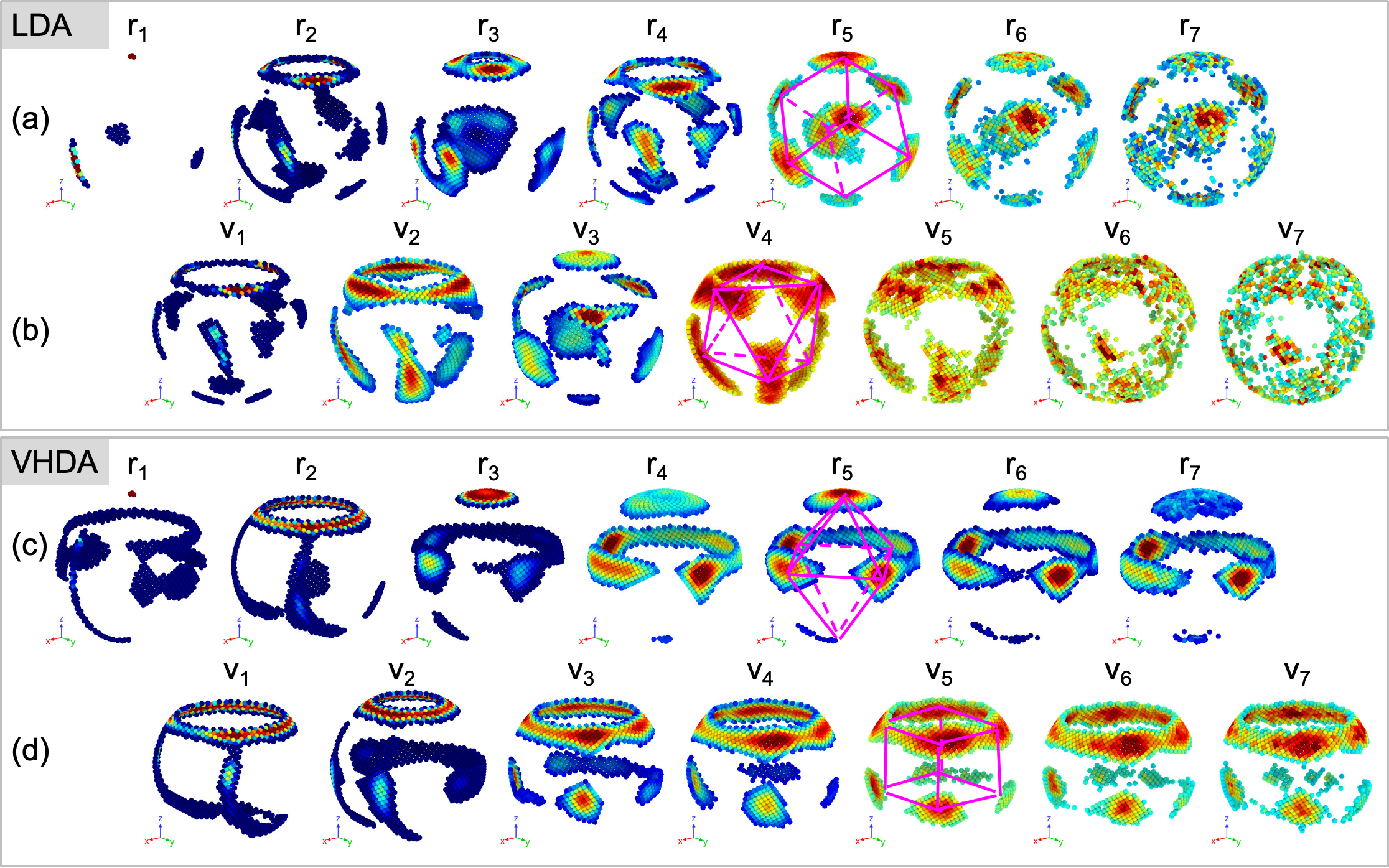}
    \caption{Evolution of three-dimensional orientational order with distance. (a)–(b) Snapshots correspond respectively to the $n$-th peak, $r_n$, and valley position, $v_n$, of $g(r)$ for LDA. (c)–(d) The same for VHDA. See Fig.~\ref{fig_2point}(a) for the  definition of these  distances. 
    To allow better visualization of the symmetry, only the part of the spherical shell that belongs to the
top 25\% of the density and covers at most 50\% of the total probability is shown. Coloring is based on the normalized density, which goes from 0.5 (dark blue) to 1 (dark red). Octahedral and cubic symmetries for the two structures at large distances are outlined by the pink lines to guide the eye. 
}
    \label{fig_3d-density}
\end{figure*}

Figure~\ref{fig_3d-density} shows for the two a-Si models the 3D density fields at the $n$-th peak and valley positions of $g(r)$, denoted as $r_n$ and $v_n$ (Fig.~\ref{fig_2point}(a)), respectively. The snapshots for $r=r_1$ show in the $z$-direction a high-density spot, a result due to the construction of the  local reference frame
with one of the bonds being along the $z$-direction.
The strip that one sees in the $x$-direction reflects the bond-angle variability, being small for LDA but quite large for VHDA, consistent with the bond-angle distribution (Fig.~\ref{fig_eos}(e)). For LDA, there are two  spots situated on the two opposite sides of the $x$-$z$ plane,  reflecting  the well-defined local tetrahedral structure. In contrast, the density distribution for VHDA is  significantly  spread out and covers non-tetrahedral regions due to the increased coordination number and the diversity of the local structure (Fig.~\ref{fig_eos}). Notably, there is an open ring around the $z-$axis that is due to a superposition  
of several strips  corresponding  to different local structures. These snapshots thus demonstrate that, at the nearest-neighbor scale, the orientational density field directly reflects the local bonding constraints.

The density distribution at $r_2$ complements the one seen at $r_1$, i.e., the high- and low-density regions are swapped. This is reasonable since the atoms in the first coordination shell ($r_1$) and the ones in the second shell ($r_2$) are not situated in the same spatial direction with respect to the local reference frame since the bond angle does not allow this. For LDA, the tetrahedral symmetry at $r_1$ gives rise to four ring-like structures at $r_2$. 
Each of these structures is composed of three substructures (strips or patches) that are due to the three atoms in the second shell that connect to the Si atom in the first shell. The density distribution at $r_2$ for VHDA is harder to discern due to mixed coordination and complex connectivity of the atoms. In fact,  around $r_2$ the structure is very varied, as can also be inferred from the small peak in $g(r)$.

Also at $r_3$ the density distributions are dual to the ones seen at $r_2$, i.e., zones of high/low density at $r_2$ correspond to low-/high-density zones at $r_3$. As a result, the distribution at $r_3$ is similar to the one found at $r_1$. For LDA, the three spots that one sees for each patch (e.g., in the $z$-direction) are due to the three (out of nine) atoms in the third shell that connect to the three atoms in the second shell. For VHDA, however, an octahedral arrangement begins to emerge at $r_3$ (with one high-density spot on the top, one at the bottom and four on the equator), despite the absence of well-defined octahedral motifs in the nearest-neighbor environment (Fig.~\ref{fig_SI_regularity-local-struc-vhda}(a)). The presence of this geometry reflects the ability of the network structure to form compact arrangements on intermediate length scales by assembling the different local motifs in a packing-efficient manner.

For LDA, the density distribution at $r_4$ is dual to the one at $r_3$ and hence resembles the one at $r_2$, indicating a well-defined sequence of rotationally shifted tetrahedral structures, similar to the arrangement found in SiO$_2$~\cite{zhang2020pnas}. 
As $r$ increases further, one observes a pronounced cubic symmetry   
at all peak positions, outlined in pink in the figure, while at the minima one finds the dual symmetry, i.e., an octahedron, see Fig.~\ref{fig_3d-density}(b). This finding can be rationalized by considering that at these distances, Si atoms start to occupy the directions of the center of the edges of the central tetrahedron, thus forming an octahedron, see the snapshot in Fig.~\ref{fig_3d-density}(b) at $v_4$. 
The next shell will thus occupy the edges of the octahedron, hence forming a  cube. These structures have higher local density than a tetrahedron and hence are preferred at large distances due to their higher packing efficiency. However, it should be noted that such an arrangement is only possible if the chains of connected Si-Si bonds that emanate from the central Si atom can bend sufficiently to access the directions corresponding to the edges of the octahedron, 
which is the case for $r> r_4$, but not for shorter distances. 
Similar results were recently found in silica glass~\cite{zhang2025pnas}, i.e., one observes the emergence of octahedral and cubic symmetries at large distances. However,  in silica one finds octahedral symmetry at the peak position of $g(r)$  and the  cubic symmetry occurs at the minimum positions of $g(r)$, i.e., the opposite of what is found  here  for amorphous silicon. This difference may be due to the  smaller angular flexibility in a-Si, which shifts the emergence of these compact arrangements to larger distances.

In contrast to  LDA, for  VHDA it is the peak positions in $g(r)$ at $r\ge r_3$ that show an octahedral symmetry while the neighboring minima in $g(r)$ (Fig.~\ref{fig_3d-density}(d)) are characterized by cubic symmetry, hence a 3D structure that is also similar to the ones found in pressurized silica glass~\cite{zhang2025pnas}.  
As mentioned above, the fact that new symmetries emerge already at shorter distances in VHDA is consistent with the densification-induced coordination changes and bond-angle distortions, which enable the network to access more compact packing configurations. Importantly, these symmetries do not originate from locally regular octahedral or cubic motifs (Fig.~\ref{fig_SI_regularity-local-struc-vhda}). Instead, they emerge only after the density field is examined over extended length scales.

From Fig.~\ref{fig_3d-density} one infers that the length scale at which the structural crossover occurs is $r^*_{\rm LDA}\approx$~8--10~\AA\ and $r^*_{\rm VHDA}\approx$~3.5--5~\AA, for LDA and VHDA, respectively.
This observation allows one to rationalize the non-trivial features in $g(r)$ pointed out earlier in the context of Fig.~\ref{fig_2point}(a). Specifically, for LDA, the splitting of the peak in $g(r)$ at $r \approx 10$~\AA\ can be attributed to
the transition from the tetrahedral structure found at small $r$ to two new 
structures (octahedral symmetry at $r \approx 8.5$~\AA\ and cubic symmetry at $r\approx 10$~\AA) that start to emerge at these length scales. 
At larger length scales, the two latter structures alternate periodically, giving rise to peaks that are spaced in a regular manner.
For VHDA, this kind of geometric transition occurs already at a shorter distance, as seen from the second peak in $g(r)$ which has an intensity close to one. At larger distances, the density field again shows alternating octahedral and cubic symmetries as discussed above. 

\begin{figure*}[ht]
\center
    \includegraphics[width=\textwidth]{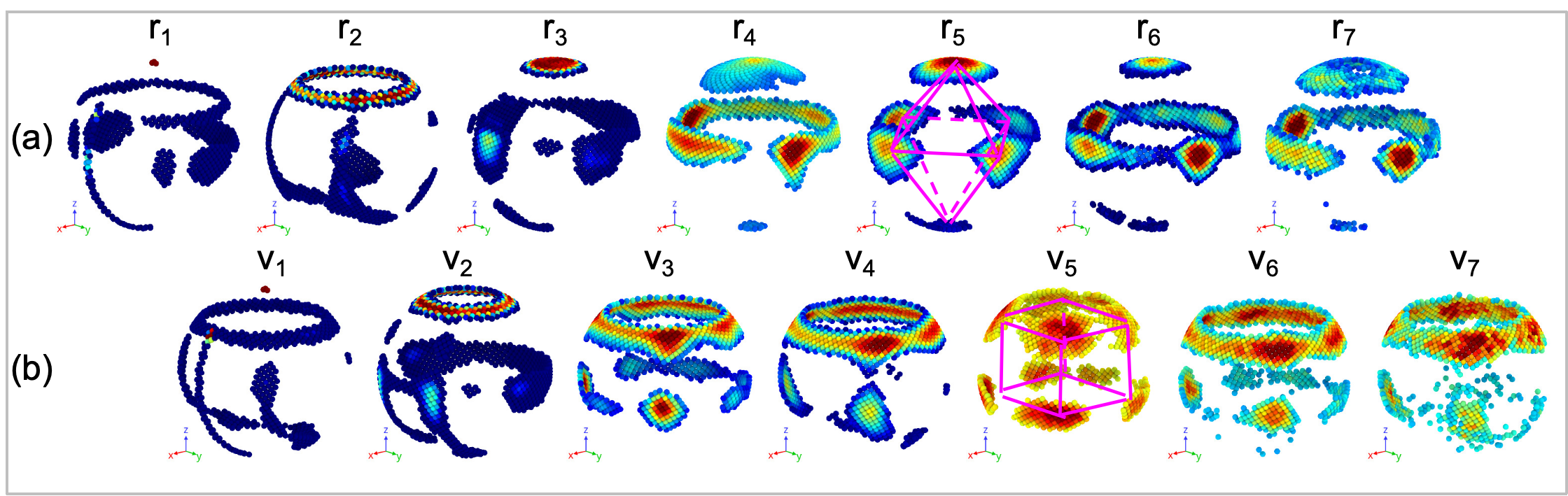}
    \caption{Evolution of the three-dimensional orientational order for the VHDA phase at 100~K. The state point (at around 16~GPa) is chosen such that this VHDA phase has a mean coordination number similar to that of the VHDA phase at 500~K.  
    }
    \label{fig_SI_3d-density-vhda-100k}
\end{figure*}

Hence, one can conclude that both LDA and VHDA show a transition of the 3D structural order from locally favored structures to packing-efficient arrangements at large length scales and it is this transition that gives rise to an unexpected feature in $g(r)$ seen in Fig.~\ref{fig_2point}(a). 
In VHDA this transition takes place at relatively small distances, since the broad bond-angle distribution (Fig.~\ref{fig_eos}(e)) allows for a large flexibility of the network, while for LDA it occurs only at intermediate distances since the narrowness of the bond-angle distribution imposes high geometrical constraints.
A very similar symmetry transition is also observed at 100~K for a VHDA phase which has the same mean coordination number as the VHDA at 500~K (Fig.~\ref{fig_SI_3d-density-vhda-100k}), indicating that this transition is primarily driven by packing constraints rather than thermal fluctuations and hence can be expected to be a generic phenomenon in glasses with a network structure.

\begin{figure*}[th]
\center
    \includegraphics[width=0.9\textwidth]{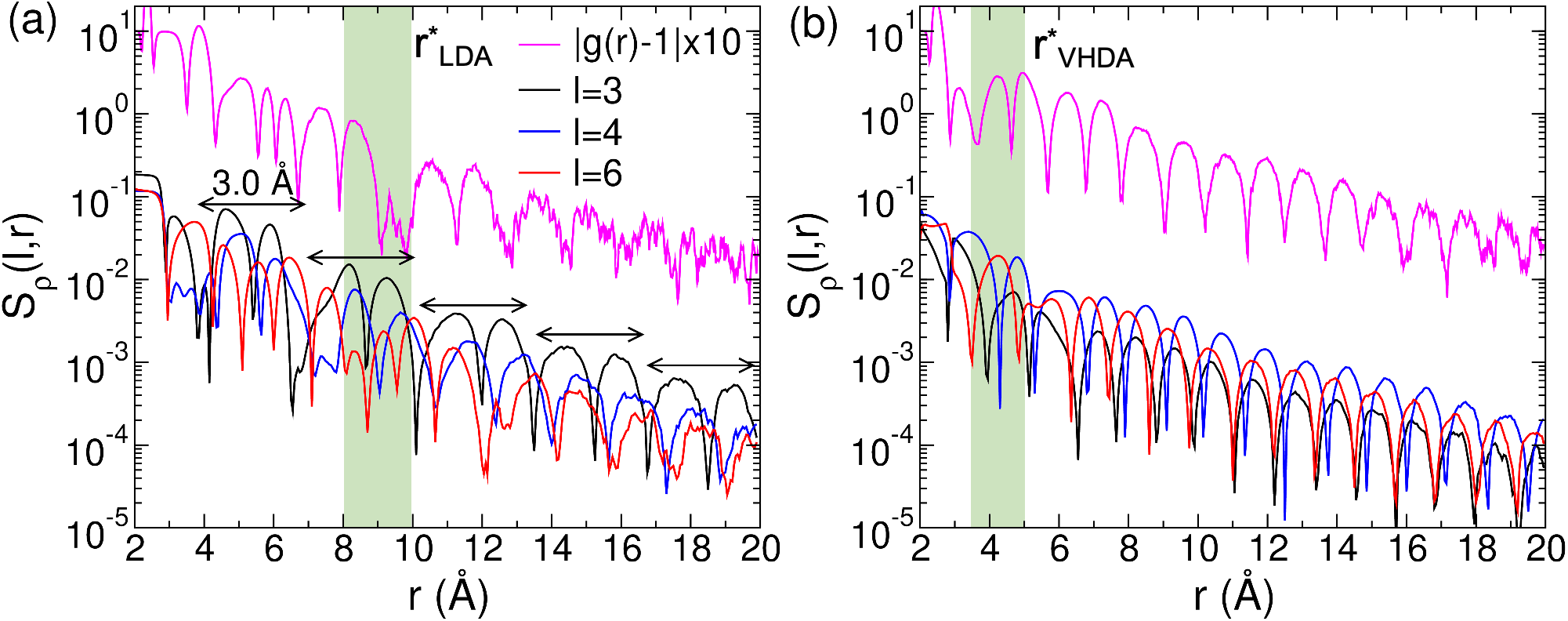}
    \caption{Angular power spectra $S_\rho(l,r)$ of the 3D density distribution for LDA (a) and VHDA (b). $|g(r)-1|$ multiplied by ten (for clarity) is included for comparison. The arrows indicate the presence of a secondary modulation with a period of around 3.0~\AA\ in $S_\rho(3,r)$ in LDA. The green vertical shaded bands mark the structural crossover zones, located at $r^*$. 
    }
    \label{fig_srho}
\end{figure*}

Further insights into the symmetry and strength of the orientational order can be obtained by expanding the density field in spherical harmonics and probing its power spectrum $S_\rho(l,r)$, see Methods for definition. Here, $l$ denotes the order (degree) of the spherical-harmonic decomposition of the density field, relevant for identifying the nature of the structural motifs in disordered systems~\cite{steinhardt1983bond,zhang2020pnas}. Figure~\ref{fig_srho} shows $S_\rho(l,r)$ for LDA and VHDA, for $l=3,4,$ and 6, i.e., the modes that are important for tetrahedral, cubic (or octahedral), and icosahedral (or dodecahedral)  motifs, respectively. For the sake of comparison, the data for $|g(r)-1|$ are also included. Note that $g(r)$ and $S_\rho$ are fundamentally different since the former quantifies the radial density fluctuations whereas the latter provides information about the symmetries of the spherical density distribution. Hence, the two quantities do not necessarily have a one-to-one correspondence in the location of their peaks. In fact, for $S_\rho(l,r)$ the peak position depends on the mode $l$, see Fig.~\ref{fig_srho}. However, one sees from  this figure that the decay of $S_\rho$ is independent of $l$ and that its slope, defined by the upper envelope of the curve, is compatible with that of $|g(r)-1|$, in accordance with previous findings~\cite{zhang2020pnas,zhang2024prb,zhang2025pnas}. 

Figure~\ref{fig_srho}(a) shows that the most pronounced mode for LDA is $l=3$, in accordance with the dominance of tetrahedral symmetry in the structure. Interestingly, the decay of $S_\rho(3,r)$ occurs in steps and the width of these steps is around 3.0~\AA, corresponding to the length scale associated with the FSDP and also to the period of the spatial arrangement of the rings, see Figs.~\ref{fig_2point}(c) and~\ref{fig_ring-para}(d). Hence, one can conclude that the FSDP in $S(q)$ reflects the spatial quasi-periodic arrangement of the ring structures, as well as the secondary modulations of the tetrahedral order in LDA Si.

From Fig.~\ref{fig_srho}(b) one recognizes that the most prominent mode for VHDA is $l=4$, in line with the observed strong octahedral/cubic symmetries. Interestingly, the secondary modulation seen in the dominant mode for LDA is absent and one observes instead a smooth regular decay of the signal. This difference is likely related to the collapse of the rings and the fact that in VHDA there is no longer a strongly dominant symmetry (like tetrahedral in the LDA), i.e., all curves have a similar value. One can also notice a strong component of $l=6$ which was found to be the dominant mode for hard-sphere-like systems~\cite{yuan2021prl}. The finding that this mode is also  pronounced in  VHDA  indicates that the system has transitioned from a tetrahedral network to a packing-dominated structural organization.

Finally, we mention that the behavior of $S_\rho(l,r)$ provides a quantitative
signature of the structural crossover. Near the crossover distance $r^*$, marked by vertical shaded bands in Fig.~\ref{fig_srho}, both $|g(r)-1|$ and the various $S_\rho(l,r)$ curves display complex, irregular features. At larger distances, by contrast, their oscillations become more regular. This confirms, beyond the visual evidence of Fig.~\ref{fig_3d-density}, that the system crosses over from local-bonding-constrained structure to a more regular, packing-driven arrangement at extended length scales.

\section{Discussion and Outlook}

Using large-scale structural models of a-Si at two densities that correspond to states with significantly different structures, we have investigated the emergence of structural organization at intermediate length scales in this prototypical network glass. Our results reveal that in a-Si the medium-range order is not simply an extension of the short-range structure. Instead, with increasing distance from a tagged particle, the network undergoes a structural crossover: Local bonding constraints dominate at short distances, whereas packing constraints generate emergent orientational organization at larger distances. This structural transition is manifested in both standard two-body and many-body descriptors of the MRO. 

The first sharp diffraction peak (FSDP) in the structure factor, $S(q)$,  has been considered for a long time to be a defining signature of MRO in  glasses. 
Previous studies have linked the FSDP in a-Si to layered structures~\cite{phillips1981topology}, crystal-like motifs~\cite{christie2004structural}, characteristic lengths associated with individual tetrahedra~\cite{shi2019distinct}, or medium-range radial correlations~\cite{dahal2021origin,dahal2022extended}. 
Our results indicate instead that the FSDP is directly connected to the spatial organization of the rings, since their characteristic inter-unit spacing coincides with the characteristic length scale of the FSDP. Thus, this finding links directly the reciprocal-space signature to the real-space organization of the network at intermediate length scales. 
Recently, similar correlations between rings in the structure and the FSDP have been demonstrated in silica-based glasses, both from neutron diffraction experiments~\cite{zhou2021experimental} and MLP-driven simulations~\cite{kobayashi2023machine}. 
This indicates that such a connection is likely to be generic in network glassy materials. The  transformation of the FSDP in LDA into the principal diffraction peak in VHDA suggests that an open network structure is required for the distinct FSDP to emerge. 
This interpretation is consistent with the observation that hard-sphere-like  systems (Lennard-Jones~\cite{binder_kob_2011}, etc.) do not have a FSDP, since such systems do not have an open-ring-like structure but are densely packed.

Beyond radial correlations, our analysis of the structure by means of four-point correlation functions reveals a transition in the symmetry character of the structure with increasing length scale, which rationalizes the anomalous features previously observed, but unexplained,  in the pair correlation functions of LDA and VHDA at $r\approx 10$~\AA\ and $r\approx 4$~\AA, respectively~\cite{durandurdu2001ab,deringer2021nature,fan2024nc}.  At short distances, the structure reflects the locally preferred organization imposed by chemical bonding and density. At large distances, however, the  prevalent symmetries are  cubic- and octahedral-like, despite the absence/scarcity (in LDA/VHDA) of such motifs in the local structure. These unexpected atomic arrangements represent  an  emergent structural organization  detectable in higher-order spatial correlations, which allow one to probe the 3D structure of the model at intermediate and large length scales. Their presence in LDA/VHDA Si, as well as amorphous silica~\cite{zhang2025pnas}, indicates that these symmetries are likely a generic feature of disordered open networks and are independent of the details of the short-range structure. Densification enhances spatial correlations and packing constraints,
thereby reducing the distance at which these unexpected symmetries occur. It is also interesting to note that the structural crossover may be observed  experimentally in the diffraction $S(q)$ in the form of a weak shoulder on the low-$q$ side of the FSDP, see inset of Fig.~\ref{fig_2point}(c).

A further key result concerns the radial distance at which the symmetries show a maximum. In LDA, cubic-like symmetries are maximized at the peaks of $g(r)$ and octahedral-like symmetries dominate at the minima, whereas in VHDA this  correspondence is reversed.
This reversal demonstrates that the distance at which the symmetry of the density field becomes octahedral (or cubic) is determined by the flexibility of the network, i.e., its ability to form a dense packing, and not by the symmetry of the structure at short distances.
This characteristic distance may also provide a structural length scale relevant for mechanical properties, such as compressibility and elasticity, a connection that should be explored in future work.

In summary, by exploiting multi-point correlation functions, we establish for the case of amorphous silicon a direct connection between the reciprocal-space signature of the MRO, the real-space organization of topological entities, and the emergence of new space-filling symmetries on extended length scales. 
More broadly, the present work proposes a transferable framework for describing the structure of network-forming glasses, in which MRO emerges from a combination of topological ring organization and local bond-angle flexibility, giving rise to complex
symmetry-selective many-body correlations. Future work exploiting this connection is promising for the design of amorphous materials with tailored functionalities.

\section*{Acknowledgements}
We thank V. L. Deringer for useful discussions. This work was supported by the National Natural Science Foundation of China (Grant No. 12474185).

\section*{Data availability}
The data and codes for the analysis supporting the findings of this study are available from the corresponding author upon reasonable request.

\clearpage

\bibliography{refs}

\end{document}